**Oxalate-precursor processing for high quality BaZrO$_3$**

Corresponding Author
**Nigel M. Kirby***
Curtin University of Technology
Department of Applied Physics
G.P.O. Box U1987, Perth 6845, Western Australia, Australia
e-mail: N.Kirby@exchange.curtin.edu.au, fax: 61 (0)8 9266 2377

Nigel M. Kirby*, Arie van Riessen, C. E. Buckley, Vaughan W. Wittorff [a]
Curtin University of Technology
Department of Applied Physics
G.P.O. Box U1987, Perth 6845, Western Australia, Australia

[a] Curtin University of Technology
Department of Electrical and Computer Engineering
G.P.O. Box U1987, Perth 6845, Western Australia, Australia
e-mail: V.Wittorff@ece.curtin.edu.au
fax: 61 (0)8 9266 2584

## Abstract

BaZrO$_3$ is by far the most inert crucible material that has been used for melt processing of high quality single crystal YBCO superconductors. To overcome the processing difficulties of existing solid-state methods, solution processing methods are increasingly important in powder synthesis. This study investigates several methods of producing oxalate precursors for subsequent thermal decomposition to BaZrO$_3$ with a view to producing high quality BaZrO$_3$ ceramics. The most favourable system used barium acetate, ammonium oxalate and zirconium oxychloride, which unlike other previously reported oxalate processes allowed near stoichiometric precipitation without requiring a large excess of Ba reagents, elevated precipitation temperatures or slow addition of reagents. Precise control over precipitate stoichiometry was achieved by variation of the solution Ba : [Zr+Hf] mole ratio without requiring accurate control over oxalate addition. XRF, XRD, N$_2$ BET adsorption, DTA/TGA and TEM analysis showed this process to be capable of producing BaZrO$_3$ powders suitable for ceramics applications. The phase purity, particle size and surface areas of BaZrO$_3$ powders produced by calcination of these precursors can be adjusted by variation of stoichiometry and calcination temperature. Crucibles formed from oxalate precursors have been able to contain Y$_2$O$_3$-BaCuO$_2$-CuO melts for up to seven days.

**Introduction**

Crucible corrosion is an important factor hindering the routine synthesis of large, high purity rare-earth barium cuprate superconductor single crystals. These compounds do not melt congruently, and must be grown in a molten flux system, for example $YBa_2Cu_3O_{7-\delta}$ is crystallised from a $BaCuO_2$-$CuO$ eutectic melt. Molten $BaCuO_2$-$CuO$ is highly corrosive to substrate materials, and the corrosion products may lead to flux and crystal contamination, poor control over crystal growth conditions, or perforation and leakage of crucibles. $BaZrO_3$ has been found to be inert to $BaCuO_2$-$CuO$ melts, but must be produced with high phase purity and low porosity to allow controlled and repeatable growth of high quality $YBa_2Cu_3O_{7-\delta}$ single crystals [1]. Off-stoichiometric or residual secondary phases lead to leakage of the crucible either through crack formation or percolation through grain boundaries. The corrosion performance of this otherwise highly resistant material is determined by trace phases, hence the key processing requirements are very accurate control over product stoichiometry and the attainment of high phase purity and density [1].

$BaZrO_3$ crucibles for melt YBCO processing have so far been produced from $BaCO_3$ and $ZrO_2$ powders by solid-state synthesis, and considerable care is needed to produce ceramics of adequate quality. To produce powders of high phase purity, repeated grinding and re-firing is necessary to sufficiently complete the solid-state reaction of $BaCO_3$ and $ZrO_2$. $BaZrO_3$ is a highly refractory material, and extended mechanical grinding is typically required to sufficiently reduce the particle size to allow sintering to high density [1]. Milling contamination from $ZrO_2$ ball milling media requires compensation by additional $BaCO_3$ and may restrict the final phase purity.

Chemically synthesized ceramic powders have numerous potential benefits over solid-state derived powders including increased phase purity, reduced particle size, reduced milling demand, and improved sintering properties. There are numerous processes reported for the production of alkali-earth zirconates and titanates, including hydrothermal, sol-gel, sol precipitation, hydroxide, peroxide, oxalate, citrate and freeze-drying processes [2-18]. Sol-gel and sol-precipitation methods in particular, have high equipment costs due to the sensitivity of reagents to moisture, and very high costs for reagents. Few investigations of solution chemical processes for crucible production for YBCO melt processing have been reported [3]. The very high level of phase purity required for melt tightness is very difficult to assess directly from bulk measurements, and hence claims of the suitability of a process to provide melt tight ceramics require verification through melt exposure.

Oxalate processing may be more suited for industrial $BaZrO_3$ processing than other chemical synthesis methods due to low equipment and reagent costs. However, based on the work of Zaitsev and Bochkarev [19,20], Potdar *et al.* [13] stated oxalate processes have certain difficulties, such as control of the speciation of solution complexes, stability of solution speciation to pH, and control of microstoichiometry. The reaction products of zirconium compounds are known to depend on zirconium solution speciation, in particular its state of hydrolysis and polymerisation [19,21]. Control over zirconium speciation adds complexity to zirconium processing and may lead to confusing or conflicting results. For example zirconium salts precipitated from freshly prepared solutions by alkali oxalates, tartrates and citrates are readily soluble in an excess of the precipitating agent, but in aged or previously boiled solutions the precipitate remains insoluble in an excess of the precipitating reagent [21]. The speciation of zirconium is dependent on factors including pH, temperature, complexing agents, solution concentration and time [21]. Because some reactions controlling solution speciation are irreversible, the complete reaction path including the order of mixing of reagents and thermal history may need to be controlled in order to direct the reaction to the desired outcome.

Previous workers have outlined oxalate processes which claim high quality $BaZrO_3$ powders can be readily achieved. Reddy and Mehrotra [14] reported a process for the production of barium zirconyl oxalate hydrate using barium chloride, zirconyl chloride and hot oxalic acid, though full details of temperatures, molar ratios of solutions, and the order and rates of addition were not reported. The precipitate was claimed to closely match $BaZrO(C_2O_4).5H_2O$ and to decompose to $BaZrO_3$ at approximately 1000ºC. The results of attempts to reproduce this experiment are provided below and are not consistent with those previously reported. Potdar et al. [13] reported a process using zirconyl nitrate and sodium oxalate to produce a soluble molecular precursor, which was subsequently reacted with barium nitrate to produce a precipitate of stoichiometric barium zirconyl oxalate. Gangadevi *et al.* [8] showed the importance of starting reagents and pH in the control of product stoichiometry. These earlier studies required either a significant excess of barium reagents, elevated temperatures or reagents containing alkalis in order to produce a stoichiometric product.

The current research was conducted to develop a simple production process yielding high quality $BaZrO_3$ powder, without requiring elevated temperatures, large excesses of barium reagents, or reagents containing alkalis. We required a powder for processing into a ceramic of sufficiently high quality for the demanding application of molten $BaCuO_2$-CuO containment. Our primary concerns were control of the stoichiometry of precursors, calcination to $BaZrO_3$ powders of high phase purity for sintering into dense ceramics of high phase purity, and verification of the tightness of sintered ceramics to YBCO melts.

## 1. Materials and Methods

### 1.1 Materials

- zirconium oxychloride, Millennium Performance Chemicals, Rockingham, Western Australia, gravimetric assay 36.8wt.% $ZrO_2+HfO_2$

- $ZrOCl_2 \cdot 8H_2O$, Riedel-de Haën 99.5%+

- $BaCl_2 \cdot 2H_2O$, AR-grade, Sigma Chemicals, Balcatta, Western Australia

- oxalic acid dihydrate, AR-grade, Sigma Chemicals, Balcatta, Western Australia

- $Ba(CH_3COO)_2$, Riedel-de Haën 99%+, gravimetric assay 99.46 wt.%

- $(NH_4)_2C_2O_4 \cdot H_2O$, Riedel-de Haën 99.5%+

### 1.2 Barium Chloride, Zirconium Oxychloride, Oxalic Acid System

Preliminary experiments were performed following the method of Reddy and Mehrohtra in which "equimolar (0.5M each) aqueous solutions of barium chloride and zirconium oxychloride were added to the hot solution of oxalic acid (1.0M) which was 10% in excess"[14]. In the current study, solutions of 0.5M $BaCl_2$, 0.5M zirconium oxychloride and 1M oxalic acid were mixed using different orders of addition at temperatures of 80 to 95ºC at solution mole ratios of 1.00:1:2.20 $BaCl_2$:[Zr+Hf]:$H_2C_2O_4$. The precipitate was stirred at constant temperature for 30 minutes, cooled to ambient temperature, filtered using Whatman #6 filter paper, washed with deionised water and dried at 120ºC. The dried precipitate was calcined in yttria-stabilised $ZrO_2$ crucibles in air at 1150ºC. A second series of experiments was performed to investigate the use of larger excesses of oxalic acid, because preliminary results showed that a severely barium deficient product was produced at solution mole ratios of 1.00:1:2.20 $BaCl_2$:Zr:$H_2C_2O_4$.

Oxalic acid addition in the barium chloride, zirconium oxychloride, oxalic acid system was optimised at a fixed 1.00:1 Ba:[Zr+Hf] solution mole ratio, at both 25º and 95ºC. A freshly prepared equimolar solution of zirconium oxychloride (0.25M) and barium chloride (0.25M) solution was added dropwise to 1M oxalic acid solution maintained at 95ºC under constant stirring. The volume of oxalic acid used was varied to study the above system with solution mole ratios of Ba:[Zr+Hf]:$C_2O_4$ over the range 1.00:1:2.50 to 1.00:1:3.00. The slurries were cooled to ambient temperature, filtered using Whatman #6 filter paper, washed twice with deionised water, dried at 100ºC in air, and calcined in air at 1000ºC for two hours.

Precipitation was also studied at 25ºC for Millennium zirconium oxychloride over the range 1.00:1:2.20-2.80 in order to assess the viability of ambient temperature production. The effect of rapid addition of mixed barium zirconium solution to oxalic acid at 95ºC at a solution Ba:[Zr+Hf]:$C_2O_4$ mole ratio of 1.00:1:2.60 was also studied.

**1.3 Barium Acetate, Zirconium Oxychloride, Ammonium Oxalate System**

The potential benefit of performing precipitation at higher pH than oxalic acid systems was investigated using barium acetate, zirconium oxychloride and ammonium oxalate over the mole ratio range of 1.00:1:2.00-3.00 respectively. Zirconium oxychloride and ammonium oxalate solutions were mixed until a clear solution was formed at 25ºC (0.075M Zr, ~0.15M $(NH_4)_2C_2O_4$), then 0.25M barium acetate was added rapidly at ambient temperature under vigorous stirring. The slurry was stirred for 90 minutes and the precipitate was filtered using Whatman #6 filter paper, washed twice in de-ionised water, dried at 100ºC then calcined in air between 1000 and 1500ºC. The same procedure was used for solution mole ratios 1.027:1:2.4 and 1.027:1:3.00.

**1.4    Ceramic Fabrication**

Powders for ceramic fabrication were produced by rapid addition of barium acetate solution to a mixed zirconium oxychloride - ammonium oxalate solution using a 1.06:1:2.50 Ba:[Zr+Hf]:$(NH_4)_2C_2O_4$ solution mole ratio at ambient temperature. Powders used for crucible fabrication were deliberately made slightly barium rich (Ba: [Zr+Hf] mole ratio 1.005 – 1.015), as our other work based primarily on solid-state derived $BaZrO_3$ has shown corrosion resistance if dramatically reduced by the presence of even trace amounts of residual $ZrO_2$ [24,25]. After washing, drying and calcination at 1300ºC, 3 wt.% cetyl alcohol was added as a pressing lubricant by ring milling in a solution containing cetyl alcohol dissolved in ethanol, after which the ethanol was evaporated at 80ºC. The lubricated powder was packed into a flexible mould with a stainless steel internal former and cold isostatically pressed at 140 MPa for 60 seconds. Sintering was conducted in a molybdenum silicide resistance furnace in air for 6 hours at 1700ºC. Sintered density was determined by Archimedes method.

Crucibles of 7mL capacity were tested for corrosion resistance to a mixture of $Y_2O_3$, $BaCO_3$ and CuO (mole ratio of 1:32:90 respectively) at 1050ºC in air. The rate of leakage was observed visually for up to seven days.

### 1.5    Analysis Methods

Millennium zirconium oxychloride, barium acetate and barium chloride reagents were standardised gravimetrically. Zirconium oxychloride was precipitated with DL-mandelic acid from acidic solution and assayed gravimetrically as ignited $Hf/ZrO_2$ [21]. The mole ratio of $HfO_2:ZrO_2$ was determined by XRF as described in earlier work [22]. In this study, Hf was assumed to have identical chemical properties as Zr, and XRF analysis was conducted for molarity control from gravimetric assays of Zr+Hf. Barium reagents were standardised by gravimetric assay of $BaSO_4$ precipitated from HCl-acidified solutions with dilute $H_2SO_4$ after ignition at 1000 ºC.

BaO : $[ZrO_2+HfO_2]$ mole ratios for all samples in this study were determined by x-ray fluorescence spectrometry (XRF) to an accuracy of ±0.002 using a procedure described in detail previously [22]. X-ray diffraction (XRD) analysis was conducted using a Siemens D500 diffractometer with a Cu tube, using 1º incidence slits, 0.15º receiving slits, a graphite secondary monochromator, and scan speed of 0.3º 2θ/min with step increment of 0.02º or 0.04º 2θ. Analysis at 40 mA and 0.3º 2θ/min with step increments of 0.04º 2θ provided a detection limit of 0.15 wt.% $BaCO_3$ (3σ counting errors) determined from calibration using experimental standards. Crystallite size was estimated by Voigt function profile fitting using the method of de Keiser *et al.* [23]. Voigt function profiles were fitted to the (100) diffraction peak of $BaZrO_3$ using SHADOW v.4.2 (Materials Data Inc. 1999). $LaB_6$ (NIST CRM 660a) was used to measure instrument broadening.

DTA/TGA analysis of precursors dried at 120ºC was conducted on a Setaram TAG24 instrument using a heating rate of 10ºC per minute to 1300ºC in air. Specimens were analysed using Pt crucibles using alumina as a reference. Multipoint $N_2$ BET analysis was conducted using an ASAP 2400 surface area analyser (Micromeritics Inc.) on samples vacuum dried at 200ºC. Powder specimens for TEM analysis were dispersed in water using ammonium polyacrylate dispersant (Dispex A-40, Allied Colloids) and dried on holey carbon grids. TEM analysis was conducted using a JEOL JEM-2011 operated at 200kV.

## 2. Results and Discussion

### 2.1 Barium Chloride, Zirconium Oxychloride, Oxalic Acid System

Powders produced in preliminary experiments using $BaCl_2$, zirconium oxychloride and oxalic acid with a solution mole ratio of 1.00:1:2.20 Ba:[Zr+Hf]:$C_2O_4$ were severely barium deficient irrespective of reaction conditions (Table 1). A 10% excess of oxalic acid was not sufficient for stoichiometric precipitation when using a 1.00 : 1 $BaCl_2$ : zirconium oxychloride solution mole ratio as has previously been claimed [8].

<Table 1 HERE>

A second series of experiments was performed to investigate the use of larger excesses of oxalic acid . because preliminary results showed that a severely barium deficient product was produced at solution mole ratios of 1.00:1:2.20 $BaCl_2$:Zr:$H_2C_2O_4$. Figure 1 shows the effect of oxalate excess on product stoichiometry for the $BaCl_2$, zirconium oxychloride and oxalic acid system using a 1.00 : 1 Ba:Zr+Hf solution mole ratio. The addition of greater than 10% excess of oxalic acid was useful in controlling product stoichiometry in the $BaCl_2$, zirconium oxychloride, oxalic acid system at both ambient and elevated temperature. In all processes investigated, the stoichiometry of the product was dependent on the initial mole ratio of the solution used for precipitation. Barium chloride, zirconium oxychloride and oxalic acid was a poor system for controlling product stoichiometry, due to the relationship between product stoichiometry, solution ratios and temperature. Unless the ratio of $H_2C_2O_4$ : Zr was approximately 2.6:1 and not 2.2:1 as implied by Reddy and Mehrohtra [14], a stoichiometric product could not be obtained at elevated temperature without a significant excess of barium or the addition of ammonia. The sensitivity of product composition to solution ratios at ambient temperature made the barium chloride, zirconium oxychloride and oxalic acid system unsuitable for high quality process control. The system was also sensitive to acid/base additions: addition of either HCl or $NH_3$ caused reduced and increased product Ba:[Zr+Hf] mole ratios, respectively.

<Figure 1 HERE>

The chemical properties of zirconium salts are affected by their processing history, for example, speciation prior to crystallization, thermal history, extent of drying etc. This is primarily due to differences in olation and oxolation [21]. In order to confirm that the chemical properties of Millennium zirconium oxychloride were not responsible for the higher oxalic acid addition required for stoichiometric precipitation than previously reported [14], reactions were also conducted using Riedel-de Haën zirconium oxychloride. Figure 1 shows there is effectively no difference between the results obtained with Millennium or Riedel-de Haën zirconium oxychloride. At near boiling temperature, the flatter region of the 95ºC curve in Figure 1 suggests the system may be capable of stable production. However, the requirement for elevated temperature control adds undesired complexity to the process.

Fresh zirconium oxychloride – oxalic acid solutions are strongly acidic at ambient temperature and heating to 95ºC and cooling back to ambient temperature lead to irreversible precipitation and pH change. For example, a freshly mixed solution containing 0.15 M zirconium oxychloride and 0.36 M oxalic acid had a pH of 0.63 ± 0.05. Heating this solution for 30 minutes at 95ºC caused the formation of a large amount of a white colloidal precipitate, and after cooling to ambient temperature the pH dropped to 0.30 ± 0.05. This precipitate did not dissolve within several weeks at ambient temperature, thus the effect of heating acid zirconium oxalate solutions appeared to be permanent. At a mole ratio of 2.60:1 $H_2C_2O_4$ : Zr, precipitation during heating began at 74ºC, and the amount of precipitate increased as the temperature was raised to 90ºC. At a mole ratio of 2.20:1 $H_2C_2O_4$ : Zr, the solution was cloudy at ambient temperature, and the quantity of precipitate increased as the slurry was heated. The precipitation of zirconium oxalates under acidic conditions at elevated temperatures makes such solutions undesirable for barium zirconate processing by inhibiting the formation of a single phase precursor. Qualitative testing showed that as the $H_2C_2O_4$ : Ba solution mole ratio increased, the barium concentrations of the supernatant solutions decreased, and the zirconium concentration increased. Under such conditions, zirconium was solubilised by an excess of oxalic acid, resulting in a zirconium deficient precipitate above a 2.60:1 $H_2C_2O_4$ : Zr solution mole ratio.

Rapid addition of reagents in the barium chloride, zirconium oxychloride, oxalic acid system at elevated temperature caused difficulties in filtration and washing. Precipitates formed by rapid reagent addition settled much more slowly and had much more severe particle losses during filtration, leading to poor repeatability of product stoichiometry. Slow rates of reagent addition were required to produce a washable product, adding undesired complexity to process control.

## 2.2 Barium Acetate, Zirconium Oxychloride, Ammonium Oxalate System

As zirconium oxychloride and ammonium oxalate solutions were mixed to form pH neutral (7.02 ± 0.05) solutions at 25ºC, an unstable precipitate formed at the contact zone of the two solutions. The precipitate rapidly re-dissolved with stirring. No precipitate formed within 10 minutes of boiling. Zirconium oxychloride / ammonium oxalate solutions were much more resistant to precipitation during heating than zirconium oxychloride / oxalic acid solutions. Slight opalescence of the solutions was observed only after cooling to ambient temperature. However, the pH dropped to 6.45 ± 0.05 indicating chemical change as a result of heating. Zirconium oxychloride / sodium oxalate mixtures formed stable clear solutions between 50ºC and 100ºC. The stability of zirconium oxychloride / oxalate solutions upon heating was clearly pH dependent, with resistance to precipitation upon heating increasing with initial pH.

The barium acetate, zirconium oxychloride, ammonium oxalate system proved to be far more suitable for stoichiometric precipitation than oxalic acid systems because the product composition could be varied and controlled by changing the Ba:Zr solution mole ratio. This could be carried out at ambient temperature with low sensitivity to excess oxalate addition above a 2.4:1 $C_2O_4$ : Ba solution mole ratio. A slight excess of barium acetate (2.7%) was required for stoichiometric precipitation in small scale experiments (Figure 2). A slightly greater Ba excess (6%) was required for stoichiometric precipitation when the volume of solution was increased to 20L to produce powder for crucible fabrication. Other than correcting for minor scale-up effects, product stoichiometry was controlled simply by the Ba:[Zr+Hf] solution ratio in the presence of a suitable excess of ammonium oxalate. For routine production of $BaZrO_3$ we used a $(NH_4)_2C_2O_4$ : barium acetate mole ratio of 2.4, though a greater excess of oxalate can be used if desired because excess ammonium oxalate does not solubilise zirconium under the conditions studied. Alternatively, near stoichiometric precursors can also be produced using a large excess of barium acetate instead of an excess of ammonium oxalate, i.e. using solution mole ratios of 2.0:1:2.0 barium acetate: zirconium oxychloride : ammonium oxalate. However the waste barium causes unnecessary disposal problems.

**<Figure 2 HERE>**

### 2.3 Calcination and Ceramic Production

Powders derived from barium acetate, zirconium oxychloride and ammonium oxalate were calcined at temperatures up to 1500 ºC although DTA/TGA analysis (Figure 3) indicated weight loss was complete by 1100 ºC. The phase purity of powders observed by XRD increased upon calcination at higher temperatures suggesting high phase purity was achieved by solid-state reaction, or that decomposition kinetics for bulk powders were sluggish compared to 50 mg specimens for DTA/TGA analysis. Near phase-pure $BaZrO_3$ was produced a temperature of 1300ºC. $BaCO_3$ and $ZrO_2$ were not detectable by XRD for the sample with Ba : Zr + Hf mole ratio of 0.986 ± 0.002 (Figure 4). However, small amounts of $BaCO_3$ (approximately 2 wt.%) were detected in calcined samples after exposure to air, particularly for samples with an excess of barium. Barium carbonate measured by quantitative XRD could be detected in powders after a few hours exposure to air, with $BaCO_3$ increasing to a maximum within approximately 24 - 48 hours. $BaCO_3$ levels are often interpreted as indicating phase impurity and hence calcination temperatures around 1300ºC were used to produce powders for crucible fabrication even though the TGA results (Figure 3) indicated weight loss was essentially complete by approximately 1100 ºC. However, our recent work has suggested BaZrO$_3$ reacts with atmospheric CO$_2$ to form BaCO$_3$ at low levels according to the phase purity and surface area of the material [25]. BaCO$_3$ was observed in BaZrO$_3$ close to phase equilibrium with either excess Ba or Zr after grinding to a fine powder only after air exposure, with BaCO$_3$ levels increasing with increased grinding time. BaCO$_3$ levels of BaZrO$_3$ powders may be caused by incomplete phase formation, surface carbonation of BaZrO$_3$ and off-stoichiometric Ba-rich phases. It is not clear that a method for separating the contributions of phase impurity and surface areas of powders to observed BaCO$_3$ levels is available. The relatively high surface areas achievable using the oxalate process make such powders prone to reaction with air after calcination.

**<Figure 3 HERE>**

**<Figure 4 HERE>**

Figure 5 shows that crystallite size increased with calcination temperature as observed by XRD. Crystallite sizes were estimated by Voigt function profile fitting of the (100) diffraction peak of BaZrO$_3$ using the method of de Keijser *et al.* [23]. These results were consistent with decreasing N$_2$-BET surface areas with calcination temperature and can be seen directly in the TEM images (Figure 6). Crystallite sizes measured by XRD were in good agreement with primary particle sizes observed by TEM. Agglomeration in powders is clearly shown in Figure 6c. The powders required de-agglomeration after calcination, which was conveniently performed by brief milling in an ethanolic solution containing cetyl alcohol used as a solid lubricant to assist isostatic pressing. The minor increase in crystallite size above 1300ºC clearly occurs by solid-state diffusion because decomposition of the oxalate is complete at approximately 1100ºC as shown by the TGA data (Figure 3). The DTA/TGA results are similar to those reported by Gangadevi *et al.* [8] and Potdar *et al.* [13]. As for other processes, the calcination temperature should be kept to the minimum required to achieve adequate phase purity, in order to provide high surface areas for solid-state densification during sintering.

**<Figure 5 HERE>**

**<Figure 6 HERE>**

The sintering properties of powders were affected by chemical purity. For example, contamination by aluminosilicates from process water of inadequate quality dramatically improved sinterability.

Powders produced using high purity water were resistant to crystallite growth under severe calcination conditions up to 1500ºC. Using high purity water is essential for melt corrosion resistance because uncontrolled contamination may cause grain boundary defects or secondary phases that are readily corroded. For example, ~30 nm barium aluminosilicate precipitates were observed at triple points in contaminated sintered samples by STEM-EDS analysis. High purity water was used for all ceramic materials made for density and corrosion studies.

**2.4 Ceramic Properties**

High density ceramics produced from oxalate derived powders had a fine grain size of approximately 5µm (Figure 7). During YBCO melt exposure at 1050ºC crucibles with a sintered density of 6.07 g/mL (97.4% theoretical density), the first sign of melt percolation through the wall section was observed after 60 hours of exposure. The rate of melt percolation was low and crucibles were still approximately half full of melt after 6 days of corrosion exposure which is sufficient time to complete high quality single crystal growth experiments. The melt viscosity did not change significantly during 6 days of corrosion exposure and remaining melt was readily decanted from the crucible.

<Figure 7 HERE>

**3. Conclusions**

The barium zirconium oxalate system may not inherently produce a strictly stoichiometric product which decomposes to phase-pure $BaZrO_3$ as readily as those of more expensive processes. Some of the experimental difficulties of oxalate processing are illustrated in this study, including effects of temperature, pH, solution ratios and types of reagents used. These difficulties were largely overcome using the barium acetate, zirconium oxychloride, ammonium oxalate system, which is shown to be a practical process using inexpensive reagents and very simple processing equipment. Using accurate control over Ba:Zr solution ratios, the process allows precise control of product stoichiometry without requiring large excesses of barium or zirconium reagents, precipitation at elevated temperature or slow addition of reagents. The precipitate can be converted to near phase-pure ultrafine $BaZrO_3$ using a brief calcination above 1300ºC, and particle sizes and surface areas can be adjusted by varying the calcination temperature. These attributes make the process a potential candidate for industrial application. Whilst many reports have claimed to have synthesised $BaZrO_3$ powders of high quality, this is the first process using a chemically derived precursor to demonstrate the capability of producing $BaZrO_3$ ceramics able to provide sustained $BaCuO_2$-CuO melt containment, for use in YBCO single crystal growth.

**4.0 Acknowledgments**

This work was supported by an Australian Postgraduate Award Scholarship and an Australian Institute of Nuclear Science and Engineering Postgraduate Research Award. We wish to acknowledge assistance provided by ANSTO Materials Division, and in particular David Cassidy for help with particle characterisation and thermal analysis. Thanks to Millennium Performance Chemicals for support of this project including supply of raw materials. Rojan Advanced Ceramics Ltd. provided CIP equipment and Michael Smirk of UWA Soil Science and Plant Nutrition assisted with XRF analysis.

**List of Table Captions**

Table 1 - Effect of reaction conditions on product stoichiometry for $BaCl_2$, zirconium oxychloride, oxalic acid system using slow addition of reagents during precipitation.

**List of Figure Captions**

Figure 1 - Effect of oxalate excess on product stoichiometry for $BaCl_2$, zirconium oxychloride and oxalic acid system using 1:1 Ba:Zr solution mole ratio. Results for Riedel-de Haën $ZrOCl_2.8H_2O$ are shown for comparison.

Figure 2 - Control of product stoichiometry in barium acetate, zirconium oxychloride, ammonium oxalate system at ambient temperature, using ammonium oxalate excess and small adjustment of Ba:[Zr+Hf] solution ratio.

Figure 3 – DTA/TGA analysis of precipitate with Ba : [Zr+ Hf] mole ratio = 1.005 ± 0.002.

Figure 4 – XRD of sample after calcination at 1300ºC for 1 hour Ba : [Zr+ Hf] for mole ratio = 0.986 ± 0.002. The expanded view shows the absence of detectable trace secondary phase peaks.

Figure 5 - Effect of calcination temperature on crystallite size and surface area of calcined powders at 0.989 ± 0.002 Ba:[Zr+Hf] mole ratio. Uncertainties are ±2σ.

Figure 6 – TEM micrographs of samples with Ba : [Zr+ Hf] for mole ratio = 0.986 ± 0.002 after calcination at a) 1350ºC for 60 minutes; scale bar = 50 nm, b) 1520ºC for 30 minutes; scale bar = 200 nm and c) 1520ºC for 30 minutes showing agglomeration of crystallites; scale bar = 100 nm.

Figure 7 - SEM micrograph of fracture surface of sintered ceramic showing fine grain size of ceramic. Scale bar = 5 µm.

**Table 1**

| Order of mixing of reagents | Precipitation Temperature (ºC) (±3ºC) | BaO:[ZrO$_2$+HfO$_2$] mole ratio (±0.002) |
|---|---|---|
| Zr added to mixture of BaCl$_2$ and oxalic acid | 95 | 0.695 |
| Zr added to mixture of Ba and oxalic acid | 80 | 0.709 |
| mixture of Ba and Zr added to oxalic acid | 95 | 0.703 |
| oxalic acid added to mixture of Ba and Zr | 90 | 0.737 |

**Figure 1**

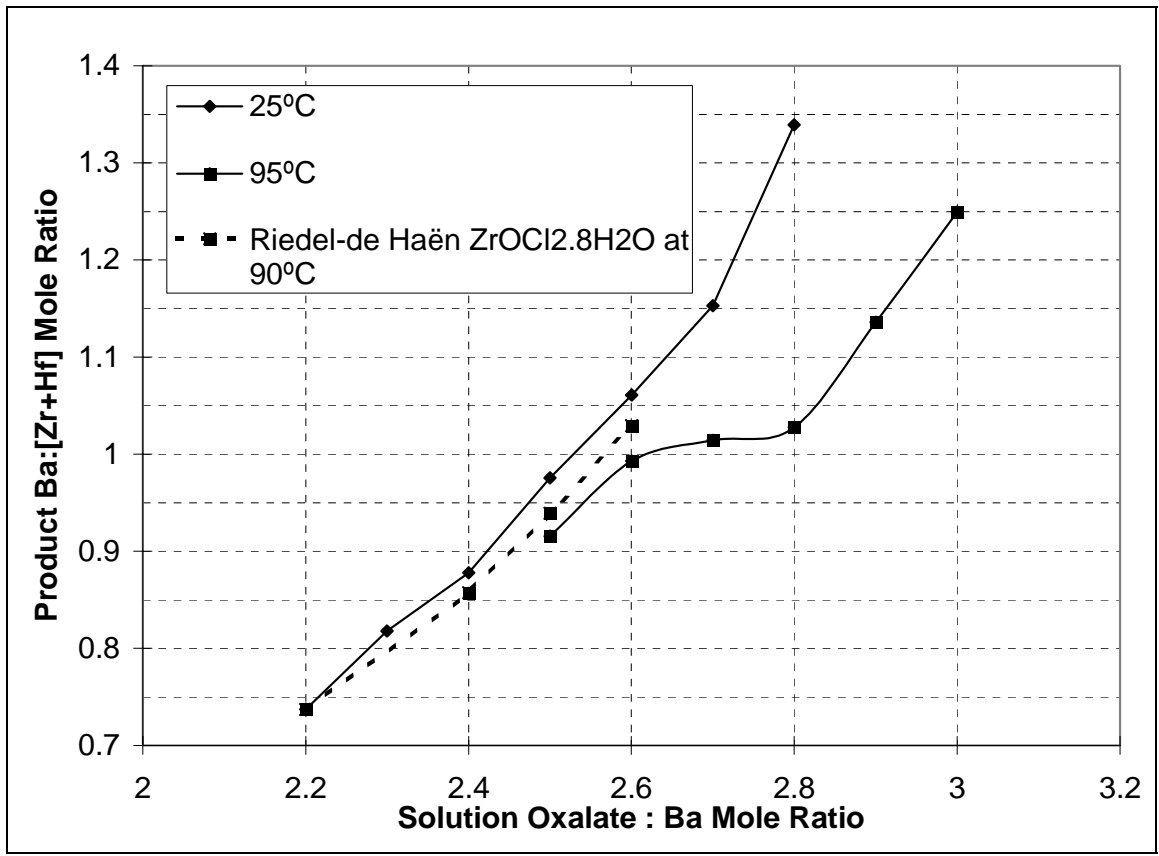

**Figure 2**

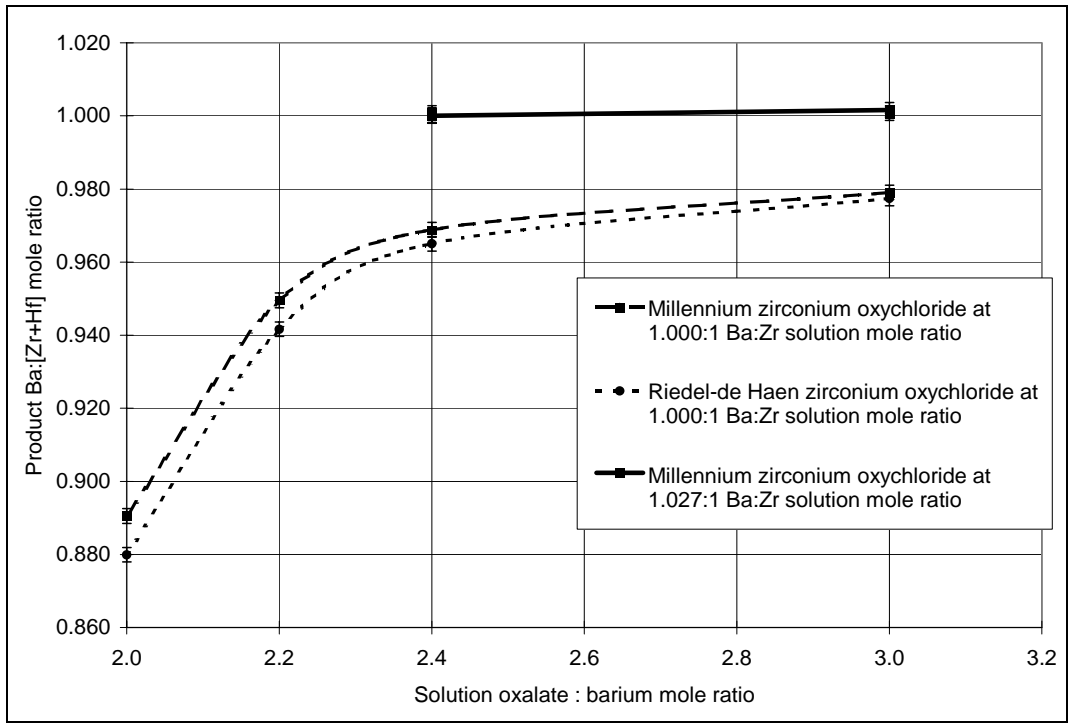

**Figure 3**

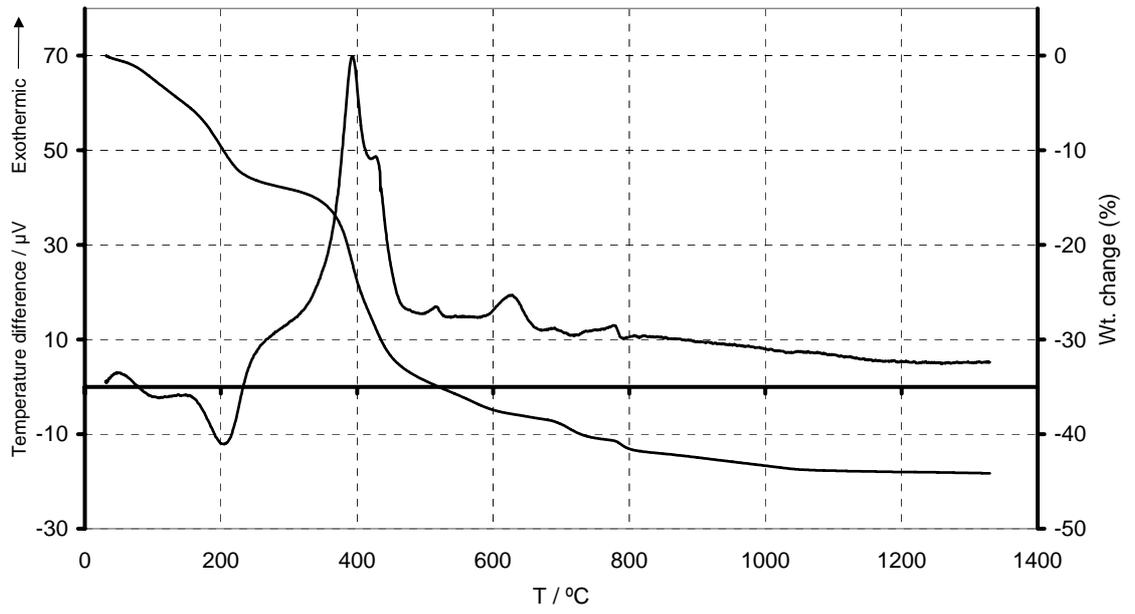

**Figure 4**

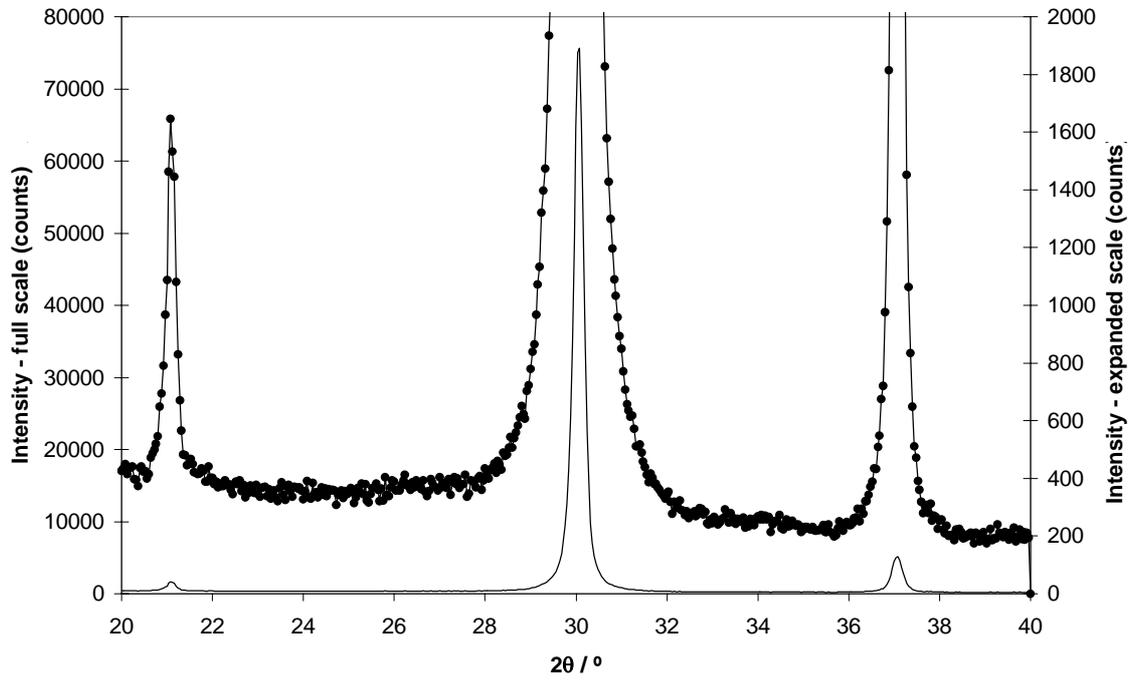

**Figure 5**

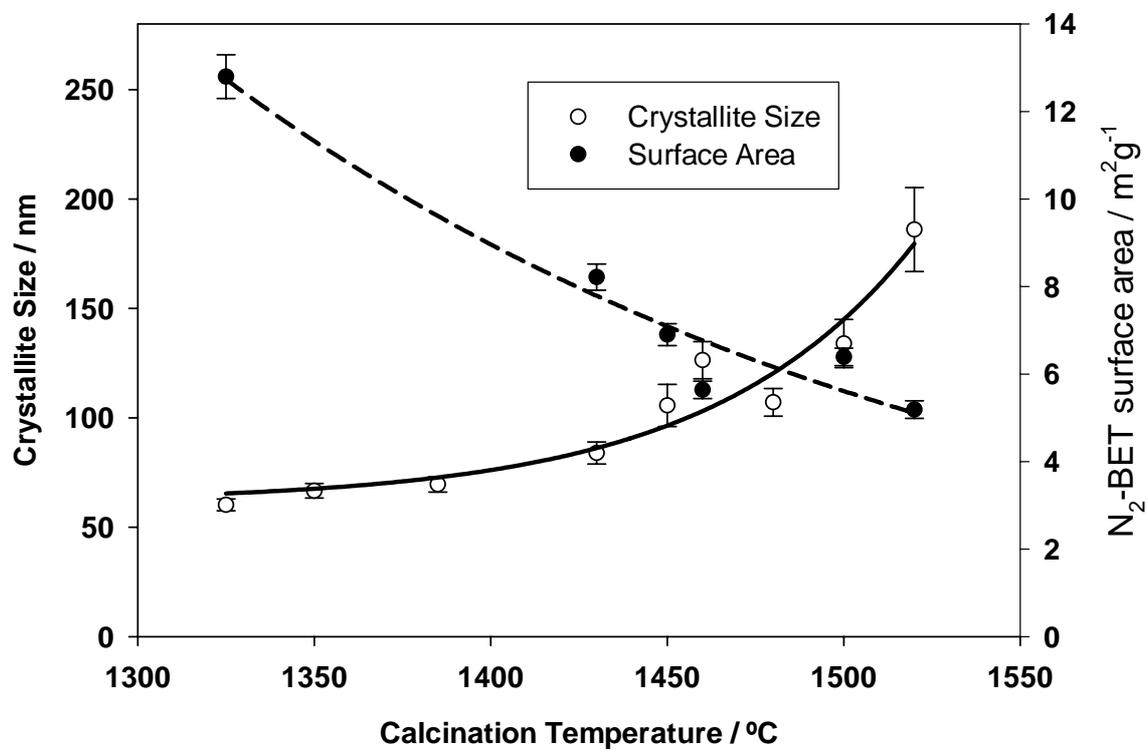

Figure 6 a

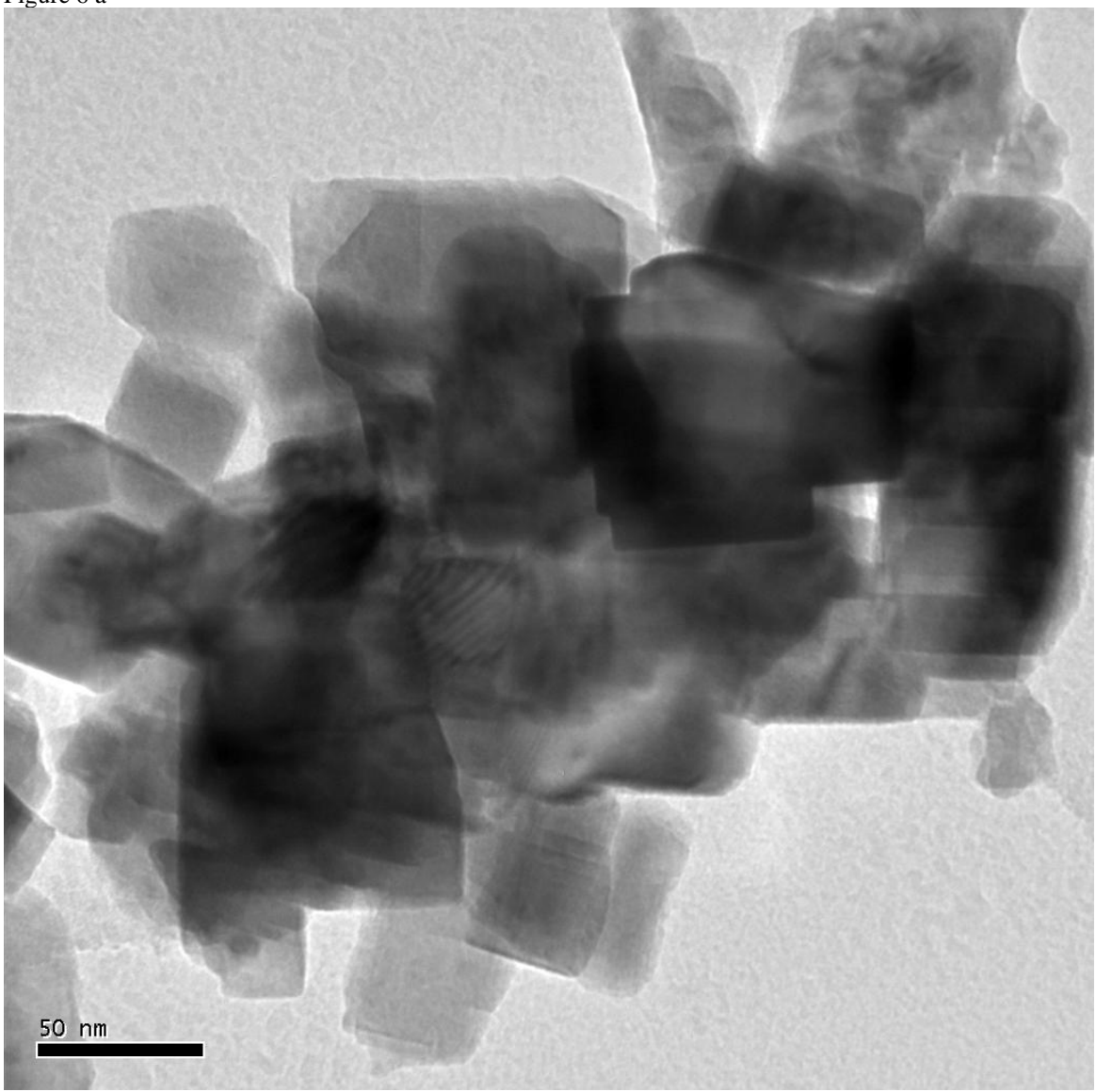

Figure 6b

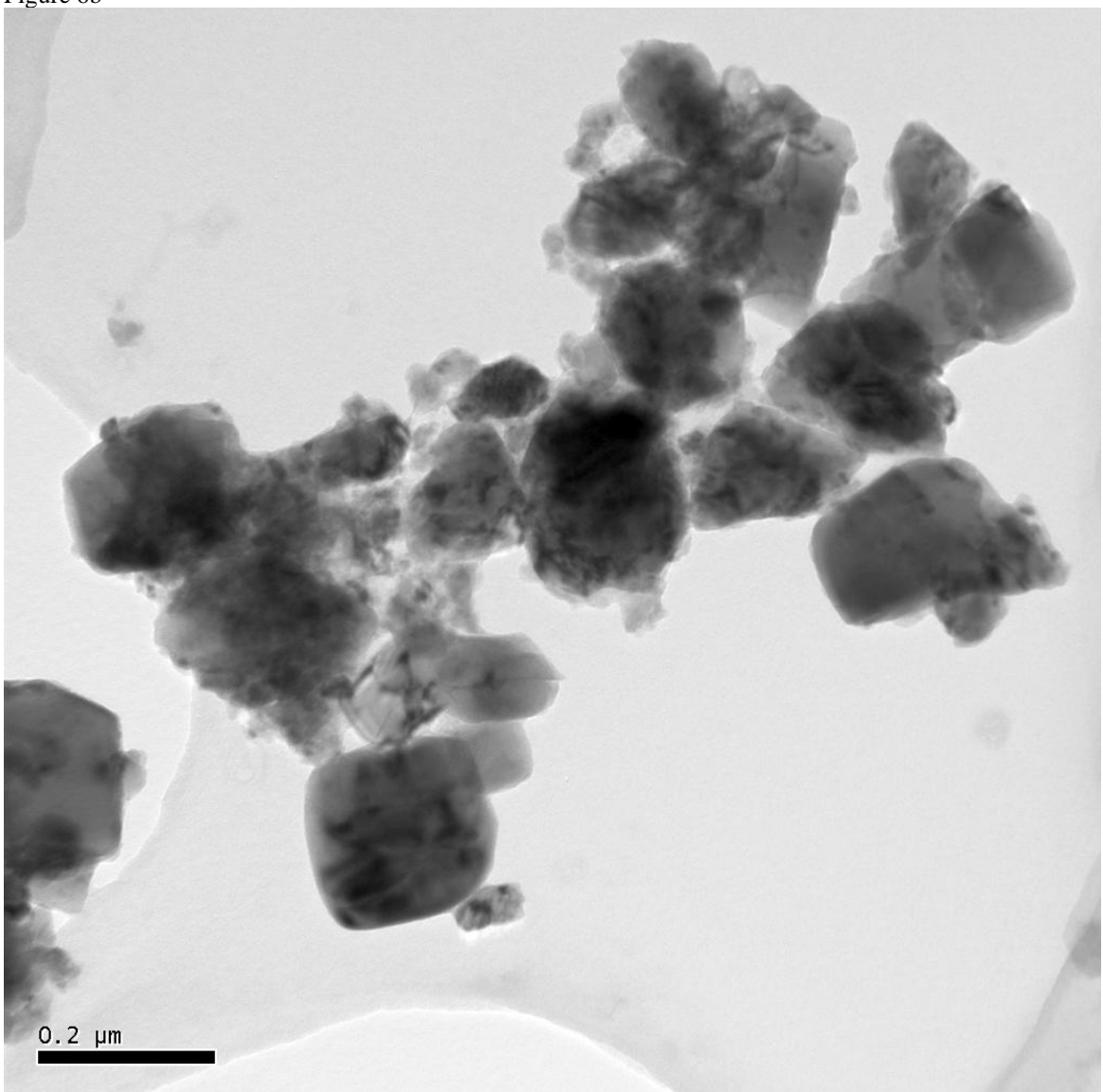

Figure 6c

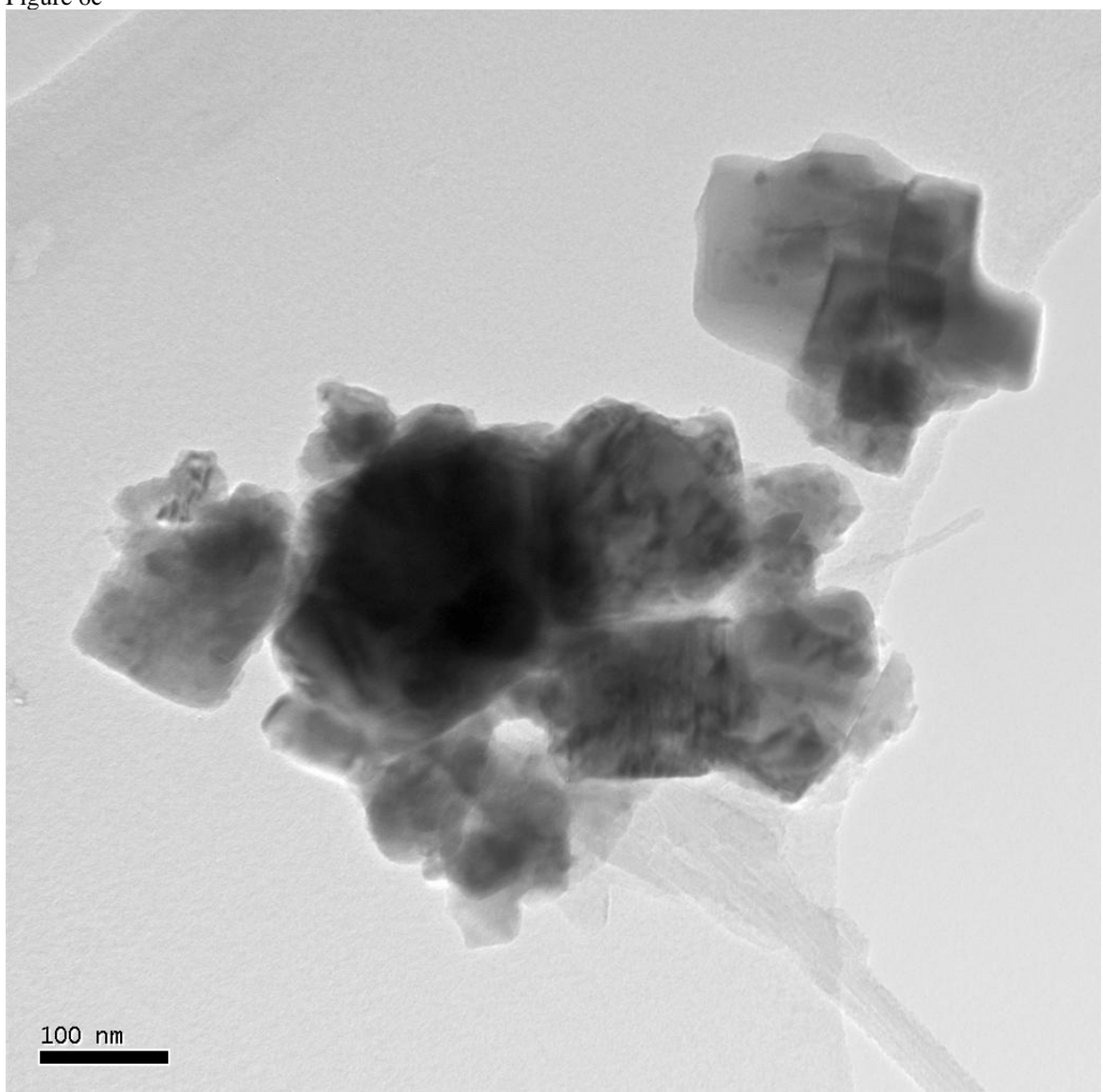

Figure 7

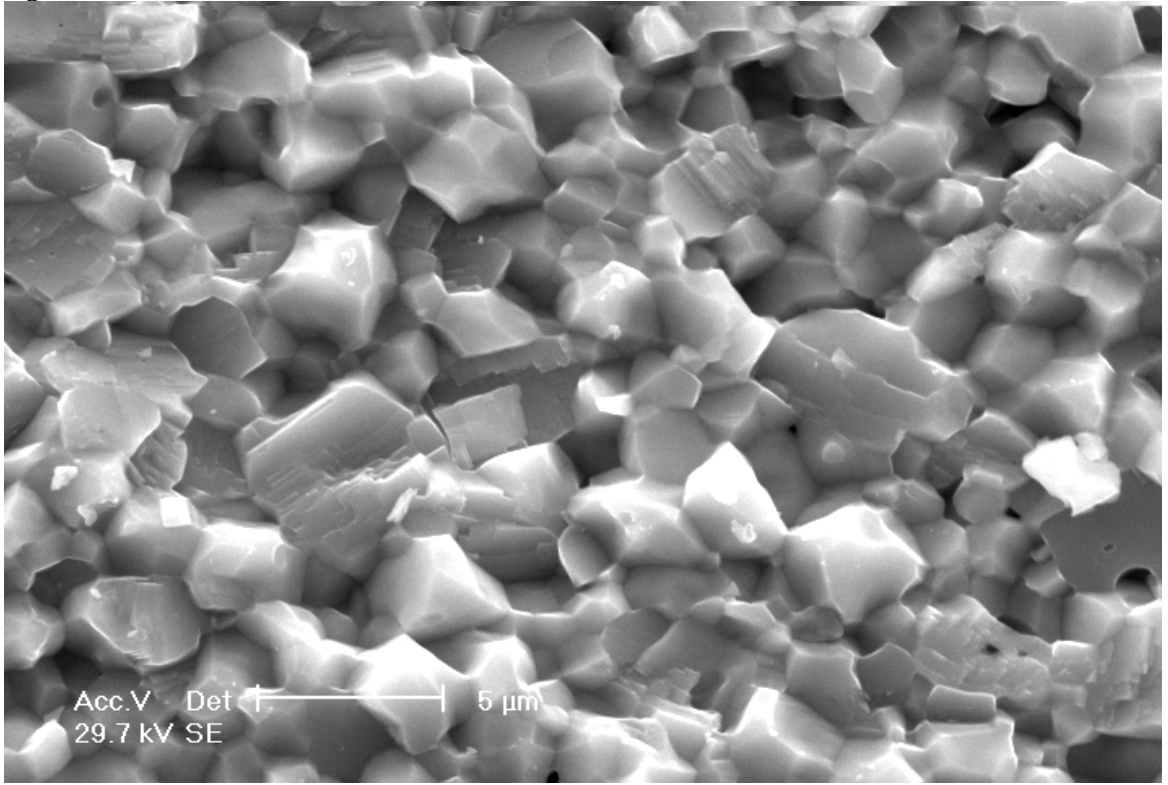